\documentstyle[prl,aps,preprint,tighten]{revtex}
\begin{document}

\draft

\title{Dynamics of 2D pancake vortices in layered superconductors}

\author{ A. S. Mel'nikov}
\address{Institute for Physics of Microstructures,
        Russian Academy of Sciences\\
         603600, Nizhny Novgorod, 46 Uljanov St., Russia\\
                  melnikov@ipm.sci-nnov.ru}

%\date{}
\maketitle
\begin{abstract}

 The dynamics of 2D pancake vortices in  Josephson-coupled
superconducting/normal - metal multilayers is considered within the
time-dependent Ginzburg-Landau theory. For temperatures close to
$T_{c}$ a viscous drag force acting on a moving 2D vortex is shown to
depend strongly on the conductivity of normal metal layers.  For a
tilted vortex line consisting of 2D vortices the equation of viscous
motion in the presence of a transport current parallel to the layers
is obtained.  The specific structure of the vortex line core leads to
a new dynamic behavior and to substantial deviations from the
Bardeen-Stephen theory.  The viscosity coefficient is found to depend
essentially on the angle $\gamma$ between the magnetic field
${\bf B}$ and the ${\bf c}$ axis normal to the layers. For field
orientations close to the layers the nonlinear effects in the vortex
motion appear even for slowly moving vortex lines (when the in-plane
transport current is much smaller than the Ginzburg-Landau critical
current).  In this nonlinear regime the viscosity coefficient depends
logarithmically on the vortex velocity $V$.

\end{abstract}
\narrowtext

\section{INTRODUCTION}

Recently the peculiarities of magnetic and transport properties of
multilayer superconducting structures attract a great deal of
attention. These investigations were stimulated, in part, by the
discovery of extremely anisotropic high-$T_{c}$ superconductors
(Bi-Sr-Ca-Cu-O, Tl-Ba-Ca-Cu-O) which may be considered as
periodic stacks of two-dimensional (2D) superconducting layers.
They pertain to a larger class of materials including quasi-2D organic
superconductors, chalcogenides and  artificial
superconducting multilayers
${YBa_{2}Cu_{3}O_{7-\delta}/(Pr_{x}Y_{1-x})Ba_{2}Cu_{3}O_{7-\delta}}$,
${DyBa_{2}Cu_{3}O_{7}/(Pr_{x}Y_{1-x})Ba_{2}Cu_{3}O_{7}}$,
$Nb/Cu$, $Nb/CuMn$, $Nb/CuGe$, $Nb/Al-AlO_{x}$
\cite{lnb00,lnb0,li,triscone,krasnov,wilson,km2}.
A common feature of these systems is the weak interlayer
Josephson coupling which results in a short effective  coherence length
$\xi_{c}$ for the order parameter spatial variation along the  ${\bf c}$
direction (perpendicular to the layers). In a broad temperature range
the $\xi_{c}$ value may be much smaller than the interlayer distance
$D$. This fact leads to the quasi-2D character of  static and
dynamic magnetic properties  in these compounds. The simplest
theoretical model taking account for such a behavior is the
Lawrence-Doniach (LD) model \cite{ld}.

Within this theory one can picture a vortex line passing through a
stack of superconducting layers (at a certain angle $\gamma$ relative
to ${\bf c}$) as a set of 2D pancake vortices connected by Josephson
vortices. A single 2D pancake vortex has its singularity (supressed
order parameter) only in one layer.
Note that in a zero external magnetic field 2D pancakes may
appear in internal superconducting layers  due to the thermally
activated nucleation of 2D vortex-antivortex pairs
\cite{minnhag,fisher}.
The interaction of 2D pancakes and the vortex lattice
structure for various magnetic field directions were considered
previously in Ref.~
\cite{minnhag,fisher,artem,clem,lnb3,lnb4,feinberg,kramer}.
 This model was also used for the description of
thermal fluctuations in the mixed state of high-$T_{c}$ superconductors
\cite{lnb1,lnb2,glazman}.

The peculiarities of viscous flux flow, flux creep and pinning in
anisotropic superconductors  are also under
intensive investigation.
In particular, the anisotropy of the flux-flow conductivity
 $\sigma_{f}$ was investigated previously  within the time-dependent
 Ginzburg-Landau (GL) theory with the anisotropic mass tensor
 \cite{asm,hao,ivlev1}. The specific feature of layered compounds
is the additional intrinsic pinning for the vortices
which are parallel to the layers.  The influence of this pinning
mechanism on the vortex motion and the peculiarities of the flux-flow
 regime for $H$ close to $H_{c2}$  were theoretically studied, for
example, in Ref.~\cite{ivlev,ivlev2}.  For weak magnetic fields
(${H_{c1}\stackrel{_<}{_\sim}H\ll H_{c2}}$) the angular dependence of
the flux-flow conductivity in highly layered superconductors was
discussed in Ref.~\cite{ivlev1}.  According to Ref.~\cite{ivlev1}, in a
tilted magnetic field in the presence of a current parallel to the
layers we have the following expression for $\sigma_{f}$:

\begin{equation}
\label{bs}
\sigma_{f}=\frac{\sigma_{0}\tilde\eta H_{c2}(\gamma=0)}
{B\mid cos\gamma\mid}
\end{equation}
where $\sigma_{0}$ is the normal-state conductivity in the plane of the
layers. Note that   $\sigma_{f}$ does not depend on the orientation of
the current with respect to the in-plane magnetic field component. The
dissipation is produced by the motion of 2D vortices formed by the
component $B_{c}$. The numerical factor   $\tilde\eta$ is determined by
the core structure of a single 2D vortex  and may be calculated
using the Bardeen-Stephen theory or within the more exact microscopic
models (see, for instance, Ref.~\cite{gorkov}). This simple picture is
certainly valid  for  multilayer systems with a negligibly small
conductivity $\sigma_{1}$ of nonsuperconducting layers
(i.e. for superconductor/insulator multilayers). The problem
becomes more complicated if we consider the vortex motion in
Josephson-coupled   superconducting/normal - metal (S/N) multilayers.
For finite conductivity $\sigma_{1}$ the spatial distribution of the
electric field generated by a moving 2D pancake changes essentially
and the additional dissipation in normal  metal layers results in
the increasing of a viscous drag force acting on a moving 2D vortex.
As a consequence the viscosity coefficient cannot be evaluated within
the simple Bardeen-Stephen model.

 In this paper we investigate the peculiarities
of 2D pancake dynamics taking into account the nonzero value of the
 interlayer normal-state conductivity.  In Sec.II for zero interlayer
Josephson coupling we obtain the equation of motion for an isolated 2D
vortex in the presence of an applied transport current ${\bf j}_{t}$.
Before discussing the specific features of the 2D vortex motion, we will
first briefly review some results concerning the motion of a vortex line
in a 3D homogeneous superconductor.  The equation for the vortex velocity
${\bf V}$ in this case has the form:

\begin{equation}
\label{gork}
\eta {\bf V}=\frac{\phi_{0}}{c} {\bf j}_{t}\times{\bf n},
\end{equation}
where ${\bf n}$ is the unit vector which points in the magnetic
field direction ($z$-direction) and
$\phi_{0}$ is the flux quantum.
For $T$ close to $T_{c}$ the quantity $\eta$ may be estimated within
the simple Bardeen-Stephen model or calculated more exactly
using the time-dependent GL theory (see, for instance,
Ref.~\cite{gorkov}).  There are two terms in the viscosity
coefficient (${\eta=\eta_{1}+\eta_{2}}$) accounting for two different
dissipation mechanisms:
 1)  the dissipation due to the relaxation of the order parameter
magnitude in the vortex core;
 2) the dissipation connected with the relaxation of the
gauge-invariant scalar potential

$$
\mu=\varphi+\frac{\hbar}{2e}\frac{\partial \theta}{\partial t}
=\varphi-\frac{\hbar}{2e}{\bf V}\nabla\theta.
$$
 $\varphi$ is the electric potential and $\theta$ -
the phase of the complex order parameter
which is assumed to be the function of the quantity
${{\bf R}={\bf r}-{\bf V}t}$, where ${{\bf r}=(x,y)}$.
The expressions for $\eta_{1}$ and $\eta_{2}$ may be written as follows:

\begin{eqnarray}
\label{eta00}
\eta_{1}=\eta_{0} \int\limits_{0}^{\infty} R
\left(\frac{\partial \rho}{\partial R}\right)^{2}
dR=0.279\eta_{0}\\
\label{eta01}
\eta_{2}=\eta_{0}\alpha=\eta_{0} \frac{1}{\pi V^{2}}
\int\rho^{2}({\bf V}\nabla\theta)
\left({\bf V}\nabla\theta-\frac{2e}{\hbar}\varphi\right)d^{2}R
=-\eta_{0}\frac{2e}{\pi\hbar V^{2}}\int
\rho^{2}({\bf V}\nabla\theta)\mu d^{2}R
\end{eqnarray}
$$
\eta_{0}=\frac{\sigma_{0} u\phi_{0} H_{c2}}{2c^{2}}
$$
where ${\rho exp(i \theta)}$
is the equilibrium order parameter
for a vortex line and $R$ - the
distance from the vortex line axis.
The function $\varphi$ satisfies the following equation
$$
\xi^{2}\Delta\varphi=u\rho^{2}
\left(\varphi -\frac{\hbar}{2e}{\bf V}\nabla\theta\right),
$$
where $\xi$ is the coherence length.
The numerical factor $u$  is determined by the pair-breaking
mechanism.  If the lack of
the energy gap is connected with the strong electron-phonon relaxation,
then $u=5.79$ (Ref.~\cite{watts}).  For superconductors with high
 concentration of magnetic impurities we have $u=12$ (Ref.~\cite{eliash})
 and ${\alpha=0.159}$ (see Ref.~\cite{kupr,hu}).

Both the dissipation mechanisms take place
for a moving 2D pancake.
The first term in the viscosity coefficient ( $\eta_{1}$)
is determined only by the spatial distribution of
 ${\rho ({\bf R})}$ in the film plane and may be calculated
in analogy with the case of a 3D vortex line (Eq.~(\ref{eta00})).
The second term ($\eta_{2}$) depends essentially on the
$\mu$  potential distribution.
For a moving 3D vortex line  $\mu$
does not depend on $z$ and decays at a characteristic
length ${\sim \xi}$ from the vortex line axis.
For a 2D pancake
moving in the $n$-th superconducting film of the
multilayer structure
the asymptotic behavior of the $\mu$
potential is essentially different.
The normal currents
in nonsuperconducting layers lead to the penetration
of the field ${\bf E}$ generated by a moving 2D vortex at a finite
length  along the $z$-axis normal to the layers and, as a consequence,
to the decreasing of $\varphi$ in the plane $z=nD$ ($D$ is the distance
between superconducting layers). It will be shown in Sec.II that
at large distances $R$ from the 2D vortex center
 ${\mu (R,z=nD)}$ decreases as  $R^{-1}$ to the distances
$$
R\sim L_{v}=\frac{8(T_{c}-T)\xi^{2}}{\pi \hbar V},
$$
where $\xi$ is the coherence length in superconducting layers.
  Such an extremely slow decay of $\mu (R)$
  results in the logarithmic divergence
of the integral in the expression (\ref{eta01})
for $\eta_{2}$ (since ${({\bf V}\nabla\theta)\sim R^{-1}}$ and
${\rho(R)\simeq 1}$ for $R\gg\xi$).
This divergence is cut off at a large length scale $L_{v}$.

These arguments show that in a 2D pancake motion equation we should
take into account not only the terms linear in $V$ but also the terms
 of the order of ${V ln(L_{v}/\xi)=V ln(V_{c}/V)}$,
 where $V_{c}=8(T_{c}-T)\xi/\pi\hbar$.
 From this point of view the situation here is similar to the one
 taking place for a vortex line in a neutral
 superfluid \cite{onuki,neu}.
The equation of motion for a 2D pancake vortex
appears to be essentially nonlinear.  The viscosity coefficient depends
strongly on the $\sigma_{1}$ value and on the vortex
velocity $V$. These results indicate a breakdown of linear response
within the employed model. However one may expect linear response to
hold for more realistic models which consider, for instance, the viscous
motion of vortex lines consisting of 2D pancakes and take into account
the interlayer Josephson coupling. In this case the logarithmic divergence
of $\eta$ may be cut off at a characteristic length which is
either the intervortex spacing or the Josephson length (see Sec.III).

 An isolated 2D vortex is surely energetically
unfavourable but the solution of this problem is important for
understanding of the peculiar dynamics of a tilted vortex line.
This dynamics is investigated in Sec.III.
 The viscosity coefficient  for a vortex line
(and hence the factor $\tilde\eta$ in
Eq.~(\ref{bs})) are shown to depend on the angle $\gamma$.  For field
 orientations close to the layers the qualitative deviations from the
 Bardeen-Stephen theory become substantial.  The nonlinear effects in
the vortex line motion are also considered.
The mechanism of these nonlinear phenomena is specific for S/N
multilayers and differs essentially from the one proposed in
Ref.~\cite{lark1,lark2} for moving flux lines in 3D homogeneous
superconductors.

 In real layered structures the interlayer Josephson coupling may be
 essential and the effect of this coupling on the 2D pancake dynamics
 should be taken into account. For rather small velocities $V$
and large angles $\gamma$ the
 logarithmic divergence of $\eta$ discussed above may be cut off at
 a length scale set by the Josephson critical current $J_{c}$.
The logarithmic dependence of $\eta(V)$ takes place above
a critical velocity determined by the $J_{c}$ value. We will
 discuss these effects qualitatively in Sec.III.

\section{  VISCOUS MOTION OF A SINGLE 2D VORTEX}

Let us consider an infinite stack of thin superconducting films of
thickness $d$ separated by  normal metal layers of thickness
${D\gg d}$.
 A good theoretical starting point for the description of static
magnetic properties of this system is known to be the LD model.
For the analysis of the dynamic phenomena (for $T$ close to $T_{c}$)
one may use  the simple generalization of this model (see, for
instance, Ref.~\cite{ivlev2})
which can be written in analogy to the time-dependent
GL model in 3D homogeneous  superconductors. It is well-known
that such an approach  corresponds to the
real situation for gapless superconductors.

In this section we will neglect the Josephson coupling of
superconducting layers (the effect of this coupling will be discussed in
Sec.III). For this particular case the general equations of the LD model are
diagonal in the layer index $n$.
The equations for the order parameter
${\rho_{n} exp(i \theta_{n})}$ and
the current density in the superconducting layer $n$ have the form:

\begin{eqnarray}
\label{main1}
\frac{\pi\hbar}{8(T_{c}-T)}\frac{\partial \rho_{n}}{\partial t}
=\xi^{2}\nabla_{\bot}^{2}
\rho_{n}+\rho_{n}-\rho_{n}^{3}-\xi^{2}\rho_{n}
\left(\nabla_{\bot} \theta_{n}-
\frac{2e}{\hbar c} {\bf A}_{\bot n}\right)^{2}\\
\label{main2}
\frac{\pi\hbar}{8(T_{c}-T)}\rho_{n}^{2}
\left(\frac{\partial \theta_{n}}{\partial t}+
\frac{2e}{\hbar}\varphi_{n}\right)=
\xi^{2}\nabla_{\bot}\left(\rho_{n}^{2}\left(\nabla_{\bot}\theta_{n}-
\frac{2e}{\hbar c}{\bf A}_{\bot n}\right)\right)\\
\label{main3}
{\bf j}_{n}=
\frac{\hbar c^{2}}{8\pi e \lambda^{2}}
\rho_{n}^{2}
\left(\nabla_{\bot} \theta_{n}-
\frac{2e}{\hbar c} {\bf A}_{\bot n}\right)
+\sigma_{0} {\bf E}_{\bot n}={\bf j}_{sn}+\sigma_{0} {\bf E}_{\bot n}
\end{eqnarray}
Here  (xy) is the film plane ,
${\varphi_{n}(x,y)=\varphi(x,y,z=nD)}$,
${{\bf A}_{\bot n}={\bf A}_{\bot}(x,y,z=nD)}$- the vector potential
component perpendicular to the $z$-axis,
${\sigma_{0}{\bf E}_{\bot n}}$-
the quasiparticle current density,
$\lambda$- the London penetration depth,
${\nabla_{\bot}=(\frac{\partial}{\partial x},\frac{\partial}{\partial
y})}$, ${{\bf E}=-\nabla \varphi
-\frac{1}{c}\frac{\partial{\bf A}}{\partial t}}$,
${{\bf E}_{\bot n}={\bf E}_{\bot}(x,y,z=nD)}$.

Note that the superconducting layers are assumed to have a negligible
thickness.
In  normal metal layers the expression
for the current is:
${{\bf j}=\sigma_{1}{\bf E}}$. Using the total current
conservation condition
 we can obtain the equation for
${\bf E}$ in nonsuperconducting layers and the boundary condition
for $E_{z}$ at $z=nD$:

\begin{eqnarray}
div{\bf E}=0\\
\frac{\sigma_{1}}{d}(E_{z}(z=nD+0)-E_{z}(z=nD-0))=- div{\bf j}_{n}
\end{eqnarray}
In the absence of an applied transport current the phase distribution
$\theta_{n}$ for a 2D pancake in the film  $n=0$ is

\begin{equation}
\nabla_{\bot}\theta_{n}=
\frac{{\bf z}_{0}\times{\bf r}_{0}}{r}\delta_{n0},
\end{equation}
where ${\bf z}_{0},{\bf r}_{0}$  are unit vectors
of the cylindrical coordinate system.
The magnetic field distribution for a 2D pancake was found in
Ref.~\cite{artem,clem}.
In the limit ${\lambda^{2}/(Dd)\gg 1}$
the expression for the vector potential ${\bf A}$  is:

\begin{equation}
{\bf A}=\frac{\hbar c}{2er}\sqrt{\frac{D}{2\Lambda}}
\left(exp\left(-\mid z\mid \sqrt{\frac{2}{\Lambda D}}\right)-
exp\left(-\sqrt{\frac{2(z^{2}+r^{2})}{\Lambda D}}\right)\right)
{\bf z}_{0}\times{\bf r}_{0},
\end{equation}
where ${\Lambda=2\lambda^{2}/d}$.
One can see that
${\bf A}$ at $z=0$
appears to be small in comparison with the value
${\frac{\hbar c}{e}\nabla \theta}$
if
${dD/\lambda^{2}\ll 1}$.
This fact results, in particular, in the logarithmic divergence
of a 2D pancake energy and in the logarithmic
interaction of 2D pancakes in the vortex-antivortex pair
 for arbitrary separations \cite{artem,clem}.
Below we will consider only the case ${dD/\lambda^{2}\ll 1}$
since this condition  may be easily fulfilled
in real superconducting superlattices at least for temperatures
close to $T_{c}$.
This approximation allows us to neglect the magnetic field generated by
 the moving 2D vortex in Eqs.~(\ref{main1})-(\ref{main3}).
For the analysis of the problem of a 2D vortex motion
  we choose the coordinate system
${{\bf R}={\bf r}-{\bf V}t}$ moving with the vortex velocity ${\bf V}$
and assume that the functions $\rho_{n}, \theta_{n},
 \varphi$ depend on ${\bf r}$ and $t$ in the following way:

\begin{equation}
\label{move}
\rho_{n}=\rho_{n}({\bf r}-{\bf V}t),\quad
\varphi=\varphi ({\bf r}-{\bf V}t,z), \quad
\theta_{n}=\theta_{nv}({\bf r}-{\bf V}t)+\theta_{n}^{t},
\end{equation}
where $\theta_{n}^{t}$  is the phase distribution corresponding to
the transport current
\begin{equation}
\label{current}
{\bf j}_{t}=
\frac{\hbar c^{2}}{8\pi e \lambda^{2}}
\left(\nabla_{\bot}\theta_{n}^{t}-\frac{2e}{\hbar c}
{\bf A}_{\bot n}^{t}\right)
=\frac{\hbar c^{2}}{8\pi e \lambda^{2}}{\bf K}_{n}
\end{equation}
The quantity $\theta_{nv}$ meets the conditions:
${\nabla_{\bot}\theta_{nv}\rightarrow 0}$ if
 ${\mid{\bf R}\mid \rightarrow \infty}$  and
\begin{equation}
\label{rotor}
rot\nabla_{\bot}\theta_{nv}=2\pi \delta_{n0}
\delta({\bf r}-{\bf V}t){\bf z}_{0}
\end{equation}
 The functions $\rho_{n}, \theta_{nv}, \varphi$ may be obtained from
 the following set of equations:

\begin{eqnarray}
\label{sys1}
-\frac{\pi\hbar}{8(T_{c}-T)}{\bf V}\nabla_{\bot}\rho_{n}
=\xi^{2}\nabla_{\bot}^{2}
\rho_{n}+\rho_{n}-\rho_{n}^{3}-\xi^{2}\rho_{n}
\left(\nabla_{\bot} \theta_{nv}+{\bf K}_{n}
\right)^{2}\\
\label{sys2}
\frac{\pi\hbar}{8(T_{c}-T)}\rho_{n}^{2}
\left(-{\bf V}\nabla_{\bot}\theta_{nv}+
\frac{2e}{\hbar}\varphi_{n}\right)=
\xi^{2}\nabla_{\bot}\left(\rho_{n}^{2}\left(\nabla_{\bot}\theta_{nv}
+{\bf K}_{n}\right)\right)\\
\label{sys3}
\frac{\sigma}{d}(\varphi^{\prime}_{z}(z=nD+0)-
\varphi^{\prime}_{z}(z=nD-0))=-\Delta \varphi_{n}+\frac{u}{\xi^{2}}
\rho_{n}^{2}\left(\varphi_{n}-\frac{\hbar}{2e}{\bf V}\nabla_{\bot}
\theta_{nv}\right),
\end{eqnarray}
 $$
\sigma=\sigma_{1}/\sigma_{0}\quad , \quad
u=\frac{\hbar c^{2}}{32\lambda^{2}\sigma_{0}(T_{c}-T)}
$$
The length ${\xi/\sqrt{u}}$ is
the electric field penetration depth.
 As was pointed out in Introduction, the coefficient $u$ may be
 calculated within  the microscopic theory
 (see Ref.~\cite{watts,eliash}).
 In normal metal layers the equation for the scalar potential
$\varphi$ is:

\begin{equation}
\label{laps}
 \Delta \varphi=0
\end{equation}
We examine at first the $\theta_{nv}$ and  $\varphi$
 distributions for ${\mid {\bf R} \mid \gg \xi}$.
In this case one can put ${\rho_{n}\simeq 1}$.
The functions ${{\bf a}_{n}=\nabla \theta_{nv}}$ and $\varphi$ may be
 obtained from the system of equations
(\ref{rotor}),(\ref{sys2})-(\ref{laps}).
Let us further introduce the 2D Fourier transform:

\begin{eqnarray}
\varphi({\bf k},z)=\int \varphi({\bf R},z)
exp(i {\bf k}\cdot{\bf R}) d^{2}R\\
{\bf a}_{n}({\bf k})=\int {\bf a}_{n}({\bf R})
exp(i {\bf k}\cdot{\bf R}) d^{2}R
\end{eqnarray}
In the Fourier representation we get the equations:

\begin{eqnarray}
-i \xi^{2}{\bf k}\cdot{\bf a}_{n}=
\frac{\pi \hbar}{8(T-T_{c})}\left(\frac{2e}{\hbar}\varphi_{n}
-{\bf V}\cdot{\bf a}_{n}\right)\\
-i {\bf k}\times{\bf a}_{n}=2\pi \delta_{n0}{\bf z}_{0}\\
\label{four1}
\frac{\partial^{2}\varphi}{\partial z^{2}}- k^{2}\varphi=0,
\qquad \mbox{for} \quad  nD<z<(n+1)D\\
\label{four2}
\frac{\sigma}{d}(\varphi^{\prime}_{z}(z=nD+0)-
\varphi^{\prime}_{z}(z=nD-0))=(k^{2}+u/\xi^{2})\varphi_{n}-
\frac{\hbar u}{2e \xi^{2}}{\bf V}\cdot{\bf a}_{n}
\end{eqnarray}
For ${nD<z<(n+1)D}$ the solution of Eq.~(\ref{four1}) has the form:

$$
\varphi=c_{n}exp(kz)+d_{n}exp(-kz)
$$
Using Eq.~(\ref{four2}) and the condition
 ${\varphi(\mid z\mid\rightarrow\infty)\rightarrow 0}$ we
may calculate the coefficients $c_{n}, d_{n}$ and obtain

\begin{equation}
\varphi_{n}=\frac{\pi \hbar u V_{c} d}{2e\sigma k\xi}
({\bf k}\cdot [{\bf V}\times{\bf z}_{0}])
\frac{sh(kD)}{sh(QD)}
({\bf k}\cdot{\bf V}-i \xi V_{c} k^{2})^{-1}
exp(-QD\mid n\mid)
\end{equation}
$$
V_{c}=\frac{8(T_{c}-T)\xi}{\pi \hbar}
$$

\begin{equation}
{\bf a}_{n}=\frac{2e\varphi_{n}{\bf k}/\hbar
-2\pi V_{c}\xi\delta_{n0}{\bf k}\times{\bf z}_{0}-
2\pi i\delta_{n0}
{\bf V}\times{\bf z}_{0}}
{{\bf k}\cdot{\bf V}-i \xi V_{c} k^{2}}
\end{equation}
The $Q$ value is determined by the equation:

\begin{equation}
ch(QD)=ch(kD)+\frac{sh(kD)d}{2k\sigma}\left(k^{2}-
\frac{i u \xi^{-1}
 V_{c}k^{2}}{{\bf k}\cdot{\bf V}-i \xi V_{c}k^{2}}
\right)
\end{equation}
Let us consider the  ${\bf k}$-domain
${L_{v}^{-1}\ll\mid{\bf k}\mid\ll\xi^{-1}}$, where
$$
L_{v}=\frac{8(T_{c}-T)\xi^{2}}{\pi \hbar V}=\frac{V_{c}\xi}{V}.
$$
This ${\bf k}$ domain correspondes to the distances
${\xi\ll\mid{\bf R}\mid\ll L_{v}}$ and appears to be rather large for
all velocities ${V\ll V_{c}}$ (we will show below that the last
condition is fulfilled for all $j_{t}$ values which are much less
than the GL critical current $j_{c}$).  In this domain in  the linear
approximation  in $V$ we get the following expressions for $\varphi$
and the phase gradient in the layer $n=0$:

\begin{eqnarray}
\label{phi0k}
\varphi_{0}=\frac{i \pi \hbar}{e}
\frac{({\bf V}\cdot[{\bf z}_{0}\times{\bf k}])}{k^{2}F(k)}\\
{\bf a}_{0}=\frac{2\pi i {\bf z}_{0}\times{\bf k}}{k^{2}}
+\frac{2\pi {\bf k}({\bf V}\cdot[{\bf z}_{0}\times{\bf k}])}
{V_{c}\xi k^{4}} \left(1-\frac{1}{F(k)}\right)
\end{eqnarray}
\begin{eqnarray}
F(k)=\left((1+k^{2}\xi^{2}/u)^{2}+4k^{2}D^{2}s^{2}
+4kDs(1+k^{2}\xi^{2}/u)cth(kD) \right)^{1/2} ;\quad
s=\frac{\xi^{2}\sigma}{udD}\nonumber
\end{eqnarray}
We consider here only the  case of small $V$ values
for which the condition
\begin{equation}
L_{v}\gg L_{m}=\max [\xi,\frac{\xi}{\sqrt{u}},D,\sqrt{s}D].
\end{equation}
is fulfilled.
 For the domain
${L_{m}\ll\mid{\bf R}\mid\ll\L_{v}}$ we have:

\begin{equation}
\label{phi0R}
\varphi_{0}\simeq\frac{\hbar}{2e}
\frac{({\bf V}\cdot [{\bf z}_{0}\times{\bf R}_{0}])}{R\sqrt{1+4s}}
\end{equation}
\begin{equation}
\label{theta0R}
\nabla_{\bot} \theta_{0}\simeq
\frac{{\bf z}_{0}\times{\bf R}_{0}}{R}+\frac{1}{2V_{c}\xi}
\left(1-\frac{1}{\sqrt{1+4s}}\right)
\left({\bf V}\times{\bf z}_{0}
ln\left(\frac{C\xi V_{c}}{RV}\right)-
{\bf R}_{0}({\bf R}_{0}\cdot [{\bf V}\times{\bf z}_{0}])\right)
\end{equation}
where $C$ is the constant of the order unity.

The asymptotic behavior of the ${\mu (R,z=0)}$
potential in the linear approximation in $V$ for ${R\gg\L_{m}}$ may
be defined from Eqs.~(\ref{phi0R}),(\ref{theta0R}).  At large R
 values ${\mu (R,z=0)}$ decreases as  $R^{-1}$ to the distances
 ${R\sim L_{v}}$ (for a nonzero $\sigma_{1}$ value).  Such an
 extremely slow decay of $\mu (R)$ is caused by the normal currents
in nonsuperconducting layers.
The field ${\bf E}$ generated by a moving 2D vortex
 penetrates at a finite
length $1/Q$ along the $z$-axis
 which results in the decreasing of
$\varphi$ in the plane $z=0$.  For example, for $s\gg 1$ (which is
possible in the vicinity of $T_{c}$ and for not very small $\sigma$
values) and ${R\gg L_{m}}$ we obtain:  ${1/Q\simeq D\sqrt{s}}$ ,
${\varphi (z=0)\ll\frac{\hbar}{2e}{\bf V}\nabla\theta_{0}}$.  The
specific behavior of $\mu (R)$ results in the logarithmic divergence
of the viscosity coefficient
 which is cut off at a large length scale $L_{v}$
(see Eq.~(\ref{eta01})).

To derive a 2D pancake motion equation we use the procedure which
is similar to the perturbation
method developed in Ref.~\cite{gorkov,neu}.
Let us consider the region
 ${R\ll min[L_{v}, L_{t}]}$ ($L_{t}$- the characteristic length scale of
the $j_{t}$ variation)
and search for the solution of Eqs.~(\ref{sys1}),(\ref{sys2})
 in the form:

\begin{equation}
\rho_{0}=f+g ;\quad \theta_{0v}=\chi+\tau
\end{equation}
where ${f exp(i \chi)}$ is the order parameter
for a static 2D pancake,
  $g$ and $\tau$- small corrections which are of the order
of $V$ and ${V ln(V_{c}/V)}$. Hereafter we neglect the higher order
terms in $V$. The equations for the corrections $g$ and $\tau$ are

\begin{eqnarray}
\label{cor1}
-\frac{\xi}{V_{c}}{\bf V}\nabla f=
\xi^{2} \Delta g+g-3gf^{2}-\xi^{2} g(\nabla \chi)^{2}-
2\xi^{2}f\nabla \chi (\nabla \tau+{\bf K}_{0})\\
\label{cor2}
f^{2}(2e\varphi_{0}/\hbar-{\bf V}\nabla\chi ) =
\xi V_{c}\nabla (f^{2}(\nabla \tau+{\bf K}_{0})+2fg\nabla\chi)
\end{eqnarray}
Note that for $V=0, {\bf K}_{0}=0, \varphi_{0}=0$ the quantities
${\chi_{p}=({\bf p}\nabla )\chi}$ É ${f_{p}=({\bf p}\nabla )f}$
(${\bf p}$- an arbitrary translation vector in the plane $(xy)$) will
solve the equations (\ref{cor1}),(\ref{cor2}) for  $g$ and $\tau$.
We next multiply Eq.~(\ref{cor1}) by $f_{p}$
and integrate over the area $S$  of the circle
${\mid {\bf R}\mid \leq R_{1}}$
where $R_{1}$ meets the condition
 ${L_{m}\ll R_{1}\ll min[L_{v}, L_{t}]}$.
After simple transformations (using Eq.~(\ref{cor2}))
one finds:

\begin{eqnarray}
\int\limits_{S} f_{p}{\bf V}\nabla f d^{2}R-
\int\limits_{S}f^{2}
\chi_{p}(2e\varphi_{0}/\hbar-{\bf V}\nabla \chi)d^{2}R=\nonumber\\
\label{circle}
=\xi V_{c}\oint\limits_{L}{\bf R}_{0}
(\nabla\chi_{p}(\tau+{\bf K}_{0}\cdot {\bf R})-
\chi_{p}(\nabla\tau+{\bf K}_{0}))dl,
\end{eqnarray}
where the circuit $L$ encloses
the area $S$.
The quantity ${\tau (R=R_{1})}$ may be taken from Eq.~(\ref{theta0R})
and the potential  $\varphi_{0}$ - from Eq.~(\ref{phi0k}).
Evaluating the integrals in Eq.~(\ref{circle}) to the logarithmic
accuracy we finally obtain:

\begin{equation}
\label{2D}
\eta_{0}\left(\beta+
ln \left(\frac{1+2s R_{m}/\xi}{1+2s}\right)+
\left(1-\frac{1}{\sqrt{1+4s}}\right)
ln\frac{L_{v}}{R_{m}}\right){\bf V}=
\frac{\phi_{0}}{c} {\bf j}_{t}\times{\bf z}_{0}
\end{equation}
$$
R_{m}= max[\xi,D] \nonumber
$$
Here the constant $\beta$ is of the order unity.
The term $\eta_{0}\beta$ is connected with the
dissipation in the domain ${R \stackrel{_<}{_\sim} \xi}$.
Note that we assumed that ${u\stackrel{_>}{_\sim}1}$ which is
consistent with the microscopic theory results for gapless
superconductors.  It is easy to see from Eq.~(\ref{2D}) that the
condition ${V_{c}/V\gg 1}$  is fulfilled for all ${j_{t}\ll j_{c}}$:

$$
\frac{V_{c}}{V}=\frac{L_{v}}{\xi} \stackrel{_>}{_\sim}
\frac{c\eta_{0}V_{c}}{\phi_{0} j_{t}}\sim\frac{j_{c}}{j_{t}}\gg 1
$$
 For ${\sigma_{1}\rightarrow 0}$ we have ${s\rightarrow 0}$,
${\beta \rightarrow \alpha+0.279}$ and
Eq.~(\ref{2D}) reduces to the usual equation
\begin{equation}
\label{usual}
\eta {\bf V}=\frac{\phi_{0}}{c} {\bf j}_{t}\times{\bf z}_{0}
\end{equation}
with the coefficient $\eta=\eta_{1}+\eta_{2}$ determined by
Eqs.~(\ref{eta00}),(\ref{eta01}).
For a nonvanishing  conductivity $\sigma_{1}$ the main contribution to
the viscosity coefficient comes from the region
${max[\xi,D]\ll R\ll L_{v}}$.
Since the parameter ${s=\xi^{2}\sigma/(udD)}$ depends on $T$ as
${(T_{c}-T)^{-1}}$ we come to the conclusion that near $T_{c}$ (if
${s\gg 1}$) the 2D pancake motion equation appears to be essentially
nonlinear.
We may neglect the constant $\beta$
and   keep only the logarithmically large terms in Eq.~(\ref{2D}).
With the temperature decreasing the logarithmic term
propotional to  ${ln(V_{c}/V)}$ will
give the dominant contribution only for small velocities $V$.

All the results obtained above are relevant only to the case of an
isolated 2D pancake. In real situation we should consider a set of such
2D vortices which form, for instance, a vortex line lattice inclined at
a certain angle $\gamma$ with respect to the $z$-axis.
It is important to note that if the distances
 between 2D pancakes are less than $L_{v}$
then we can not use directly  Eq.~(\ref{2D})  for each 2D vortex.
The thing is that neighbouring 2D pancakes may change essentially the
$\mu$ potential distribution in the vicinity of the given
2D pancake ( for ${R \stackrel{_<}{_\sim} L_{v}}$).
  Therefore, the dissipation
in such a system  may depend on the arrangement of 2D pancakes.
%%%%%%%%%%%%%%%%%%%%%%%%%%%%%%%%%%%%%%%%%%%%%

\section{VISCOUS MOTION OF A VORTEX LINE}

In this section we study the motion of a single tilted vortex line
in the presence of an applied transport current parallel to the
layers.  At first let us discuss briefly the case of a static vortex
line which lies in the (xz) plane, contains the origin and forms the
angle $\gamma$ with the $z$ axis. In the layer $n$ the 2D vortex
center is at ${{\bf r}_{n}=(nDtg\gamma,0)}$. Therefore, we have a
discrete sequence of phase singularities situated in the layers:

\begin{equation}
\label{rotor2}
rot\nabla\theta_{n} = 2\pi\delta ({\bf r}-{\bf r}_{n}) {\bf z}_{0}
\end{equation}

The angle $\gamma$ depends on the magnitude and direction
of the applied magnetic field
(${{\bf H}={\bf z}_{0}Hcos\tilde\gamma+{\bf x}_{0}Hsin\tilde\gamma}$),
which determine the energetically favourable vortex structure.
In the limit of zero interlayer Josephson coupling the field component
parallel to the layers is unscreened and penetrates the structure
unpertubed, while the perpendicular component creates a vortex lattice
normal to the layers
(the so-called ``decomposition property'' \cite{feinberg}).
As a consequence, tilted vortices cannot be formed in a tilted external
field if the interlayer Josephson critical current  is zero.
It is, therefore, very important to understand the effect of a   small
Josephson coupling. For a   static vortex lattice this problem has been
previously treated within the LD theory
 (Ref.~\cite{lnb3,feinberg,kramer}).
The tilted vortex lattice was shown to be energetically favourable for
a wide domain of $H$ and $\tilde\gamma$. For the field orientation
close to the layer direction this tilted lattice may transform into the
lattice parallel to the layers
(the orientational lock-in transition \cite{lockin})
or into a new type of  vortex arrangement  which consists   of sets of
coexisting parallel and perpendicular vortices \cite{lnb3}.

The core structure for a tilted vortex line was examined in detail in
Ref.~\cite{lnb3,feinberg}.
 It was found that this structure depends essentially
on the ratio ${Dtg\gamma/L_{j}}$ where
$$
L_{j}=\sqrt{\frac{d\hbar c^{2}}{8\pi e \lambda^{2}J_{c}}}
$$
 is the Josephson length.
The case ${tg\gamma > L_{j}/D}$ corresponds to the formation of
Josephson-like vortex cores connecting 2D pancakes.  Hereafter we will
consider mainly the other angular domain ${tg\gamma < L_{j}/D}$.
In this case the in-plane currents are much larger than the Josephson
currents. This fact provides a possibility to neglect the Josephson
coupling in calculations of the vortex viscosity $\eta$.

As was pointed out in Sec.II, the dissipation produced by
moving 2D pancakes may depend on their arrangement.
The validity of this assumption
 will be demonstrated in this section for the particular case
of a tilted flux line. For rather small velocities the logarithmic
divergency of the viscosity coefficient is cut off at a length scale
determined by  the size of the so-called
effective core \cite{feinberg} (i.e. at the length ${Dtg\gamma}$).
One can say that the main contribution to the dissipation
for magnetic field orientations close to the layers comes
exactly from the effective core region.
In this paper we restrict
ourselves to weak magnetic fields and consider the case when the
effective cores in the tilted vortex lattice do not overlap.  Such an
approach is valid if ${D tg\gamma\ll a_{x}}$ where $a_{x}$ is the
distance in the $x$-direction between 2D vortices in
the same plane but pertaining to neighbouring vortex lines. It was
shown in Ref.~\cite{feinberg} that this condition is fulfilled for

$$
H<\frac{\phi_{0}}{\pi D^{2}tg\tilde\gamma};
 \quad H\ll H_{c2}
$$

 For a moving vortex line we assume the following dependence of the
quantities  $\rho_{n}, \theta_{n},
 \varphi$  on the variables ${\bf r}$ and $t$:
\begin{equation}
\rho_{n}=\rho ({\bf R}-{\bf r}_{n}),\quad
\varphi=\varphi ({\bf R},z), \quad
\theta_{n}=\theta_{nv}+\theta_{n}^{t}=
\theta_{v}({\bf R}-{\bf r}_{n})+\theta_{n}^{t},
\end{equation}
Here ${{\bf R}={\bf r}-{\bf V} t}$ and
the quantity  $\theta_{n}^{t}$ correspondes to the transport
current (\ref{current}).   We will examine at first the behavior of
the functions $\theta_{v}$ and $\varphi$ in the  domain
${\mid{\bf R}-{\bf r}_{n}\mid\gg\xi}$ (${\rho_{n}\simeq 1}$).
At the same time we
consider only the regions
${\mid{\bf R}-{\bf r}_{n}\mid\ll \lambda}$ which gives us a
possibility to neglect the vector potential ${\bf A}$. Let us
 introduce the function ${{\bf a}=\nabla\theta_{v}}$.  Using the 2D
Fourier transform

$$
\varphi({\bf k},z)=\int \varphi({\bf R},z)
exp(i {\bf k}\cdot{\bf R}) d^{2}R
$$
$$
{\bf a}({\bf k})=
\int {\bf a}({\bf R})
exp(i {\bf k}\cdot{\bf R}) d^{2}R
$$
and Eqs.~(\ref{sys2})-(\ref{laps}),(\ref{rotor2}) we obtain:

\begin{eqnarray}
-i \xi^{2}({\bf k}\cdot{\bf a})exp(i k_{x}nDtg\gamma)=
\frac{\pi \hbar}{8(T-T_{c})}\left(\frac{2e}{\hbar}\varphi_{n}
-({\bf V}\cdot{\bf a}) exp(i k_{x}nDtg\gamma) \right)\\
-i {\bf k}\times{\bf a}=2\pi {\bf z}_{0} \\
\frac{\partial^{2}\varphi}{\partial z^{2}}- k^{2}\varphi=0,
\qquad \mbox{for} \quad  nD<z<(n+1)D\\
\frac{\sigma}{d}(\varphi^{\prime}_{z}(z=nD+0)-
\varphi^{\prime}_{z}(z=nD-0))=
(k^{2}+u/\xi^{2})\varphi_{n}-
\frac{\hbar u}{2e \xi^{2}}({\bf V}\cdot{\bf a})
exp(i k_{x}nDtg\gamma)
\end{eqnarray}
After the solution of this system we have the following expressions for
the phase gradient ${\bf a}$ and the $\varphi$ potential in the plane
$z=0$:

\begin{eqnarray}
\label{aline}
{\bf a}({\bf k})=\frac{2\pi P(k) {\bf V}\times{\bf z}_{0}-
2\pi i V_{c}\xi {\bf k}\times{\bf z}_{0}}
{V_{c}\xi k^{2}+i P(k) {\bf k}\cdot{\bf V}}\\
\label{philine}
\varphi ({\bf k},z=0)= \frac{i \pi \hbar}{e}(P(k)-1)
\frac{({\bf V}\cdot [{\bf k}\times{\bf z}_{0}])}
{k^{2}+i P(k){\bf k}\cdot{\bf V}/(V_{c}\xi)}\\
P(k)=1 - \left(1+\frac{k^{2}\xi^{2}}{u}+
2s\frac{kD}{sh(kD)}(ch(kD)-cos(k_{x} D tg\gamma))\right)^{-1}
\nonumber
\end{eqnarray}
Let us now examine the case of small velocities V
for which the condition

$$
L_{v} \gg max[\xi, \frac{\xi}{\sqrt{u}}, D, D\sqrt{s}, D tg\gamma]
$$
is fulfilled.
To solve Eqs.~(\ref{sys1}),(\ref{sys2}) for the layer $n=0$  we
use  the perturbation method considered in Sec.II.
After the similar procedure one obtains the equation
(\ref{circle}) where the
radius $R_{1}$ meets the condition
$$
max[\xi, \frac{\xi}{\sqrt{u}}, D, D\sqrt{s}, D tg\gamma]
\ll R_{1}\ll L_{v}.
$$
Using Eqs.~(\ref{aline}),(\ref{philine})
 we may evaluate  the integrals in
Eq.~(\ref{circle}) taking
into account only the first order corrections in $V$. One finally has
the equation of the vortex line motion in the form (\ref{usual}) where
the viscosity $\eta$ is given by the expression:

\begin{equation}
\label{etaint}
\eta\simeq\eta_{0}\left(\beta_{0}+ \int\limits_{0}^{1/\xi}
\frac{P(k) d^{2}k}{2\pi k^{2}}\right)
\end{equation}
Here $\beta_{0}$ is the constant of the order unity
which is determined by the dissipation processes in the regions
${\mid {\bf R}-{\bf r}_{n}\mid \stackrel{_<}{_\sim} \xi}$.
We evaluated the integral in Eq.~(\ref{etaint}) to
the logarithmic accuracy (i.e. for the case when the main contribution
to $\eta$ comes from the domain ${\mid {\bf R}-{\bf r}_{n}\mid>\xi}$)
and obtained:

\begin{eqnarray}
1.D\ll\xi, Dtg\gamma\ll\xi, u\stackrel{_>}{_\sim}1 \nonumber\\
\label{eta1}
\eta\simeq\eta_{0}\left(\tilde\beta+
ln\left(\sqrt{1+\frac{\sigma D}{ud cos^{2}\gamma}}+
\sqrt{1+\frac{\sigma D}{ud}}\right) \right)\\
2.D\ll\xi, Dtg\gamma\gg\xi, u\stackrel{_>}{_\sim}1 \nonumber\\
\label{eta2}
\eta\simeq\eta_{0}\left(\tilde\beta+
\left(1-\frac{1}{\sqrt{1+4s}}\right) ln(Dtg\gamma/\xi)
+ ln\sqrt{1+s}\right)\\
3.D\gg\xi, tg\gamma\ll 1, u\stackrel{_>}{_\sim}1 \nonumber\\
\label{eta3}
\eta\simeq\eta_{0}\left(\tilde\beta+
ln \left(\frac{(1+2sD/\xi)\sqrt{1+s}}{1+2s}\right)\right)\\
4.D\gg\xi, tg\gamma\gg 1, u\stackrel{_>}{_\sim}1 \nonumber\\
\label{eta4}
\eta\simeq\eta_{0}
\left(\tilde\beta+
ln \left(\frac{(1+2sD/\xi)\sqrt{1+s}}{1+2s}\right)+
\left(1-\frac{1}{\sqrt{1+4s}}\right) ln(tg\gamma)\right)
\end{eqnarray}
The quantity $\tilde\beta$ is of the order unity.
These results show that for ${L_{v}\gg Dtg\gamma\gg max[D,\xi]}$
 the logarithmic
divergence of the viscosity $\eta$ is cut off at the length
${Dtg\gamma}$, i.e. at the distance between neighbouring 2D pancakes.
This fact results in the angular dependence of the vortex line
viscosity.

 It is natural to suppose that for the case ${L_{v}<Dtg\gamma}$
the main contribution to the dissipation produced by each 2D pancake
moving in the n-th layer comes from the distances ${\mid
{\bf R}-{\bf r}_{n}\mid \stackrel{_<}{_\sim} L_{v}}$.  As a result the
viscosity of a tilted vortex line will depend logarithmically on $V$
(see also Sec.II).  This supposition is proved by  direct
calculations. For the domain ${L_{v}^{-1} \ll k \ll \xi^{-1}}$ within
 the linear approximation in $V$ Eqs.~(\ref{aline}),(\ref{philine})
 may be written as follows:

\begin{eqnarray}
{\bf a}({\bf k})\simeq \frac{2\pi i {\bf z}_{0}\times{\bf k}}
{k^{2}}+\frac{2\pi P(k){\bf k}
({\bf V}\cdot[{\bf z}_{0}\times{\bf k}])}
{V_{c}\xi k^{4}}\\
\varphi({\bf k},z=0)\simeq (1-P(k))
\frac{i \pi\hbar ({\bf V}\cdot [{\bf z}_{0}\times{\bf k}])}
{e k^{2}}
\end{eqnarray}

Let us now assume that ${L_{m}\ll L_{v}\ll Dtg\gamma}$.
Then for the region ${L_{m}\ll R\ll L_{v}}$ one obtains the
expressions for ${{\bf a}({\bf R})}$ and ${\varphi({\bf R},z=0)}$
which coincide with Eqs.~(\ref{phi0R}),(\ref{theta0R}). To derive the
vortex motion equation we should perform the procedure described in
Sec.II choosing the $R_{1}$ value according to the condition
${L_{m}\ll R_{1}\ll L_{v}}$.  The final equation   for ${\bf V}$
coincides with Eq.~(\ref{2D}). The condition ${L_{v}\ll Dtg\gamma}$
is fulfilled for vortex velocities
${V\gg V^{*}=V_{c}\xi/(Dtg\gamma)}$ which correspond to in-plane
 transport current densities ${j_{t}\gg j^{*}}$ where

\begin{equation}
j^{*}=\frac{j_{c}\xi}{Dtg\gamma}
\left(\tilde\beta+
ln \left(\frac{(1+2sR_{m}/\xi)\sqrt{1+s}}{1+2s}\right)+
\left(1-\frac{1}{\sqrt{1+4s}}\right)
ln\left(\frac{Dtg\gamma}{R_{m}}\right)\right)
\end{equation}

Note that the results obtained above are valid only for the case
${j_{t}\ll j_{c}}$.
In the angular interval ${tg\gamma\gg\xi/D}$ the current
density $j^{*}$ may be much less than $j_{c}$ and
there exists a current domain ${j^{*}\ll j_{t}\ll j_{c}}$ where
the vortex line viscosity depends logarithmically on $V$.

Taking into account the expression for the electric field
$
\overline{{\bf E}}=\frac{1}{c}{\bf V}\times{\bf B}
$
one obtains
\begin{equation}
\label{vax}
\overline{{\bf E}}=\frac{2B\eta_{0}}{\sigma_{0}u H_{c2}\eta}
({\bf j}_{t}({\bf n}\cdot{\bf z}_{0})-
{\bf z}_{0}({\bf n}\cdot{\bf j}_{t}))
\end{equation}
where ${\bf n}$ is the unit vector parallel to ${\bf B}$, $H_{c2}$ -
the upper critical field for ${{\bf H}\parallel {\bf z}_{0}}$.
The flux-flow conductivity $\sigma_{f}$ may be written as follows:

\begin{equation}
\label{fluxflow}
\sigma_{f}=\frac{\sigma_{0}u H_{c2}\eta}
{2B\eta_{0}\mid cos\gamma\mid}
\end{equation}
 The quantity $\eta$ in Eqs.~(\ref{vax}),(\ref{fluxflow})
 is given by the expressions
 (\ref{eta1})-(\ref{eta4}) if
${j_{t}<j^{*}}$. For large angles $\gamma$ (${tg\gamma\gg \xi/D}$)
and for current densities ${j_{t}>j^{*}}$  we have

\begin{equation}
\eta=\eta_{0}
\left(\tilde\beta+
ln \left(\frac{1+2sR_{m}/\xi}{1+2s}\right)+
\left(1-\frac{1}{\sqrt{1+4s}}\right)
ln\left(\frac{\xi V_{c}Bcos\gamma}
{R_{m} c\mid\overline{{\bf E}}_{\bot}\mid}\right)\right)
\end{equation}
where $\overline{{\bf E}}_{\bot}$
is the electric field component perpendicular to
the $z$ axis.
The current-voltage characteristics in this case appear
to be nonlinear at least for temperatures close to $T_{c}$
(${s\stackrel{_>}{_\sim}1}$).
Note that for negligibly small  $\sigma_{1}$  values
the expression for $\sigma_{f}$ (\ref{fluxflow}) coincides with
Eq.~(\ref{bs}).

As was mentioned above, all these  results are valid only in the
 angular interval ${tg\gamma < L_{j}/D}$.
In the opposite case ${Dtg\gamma > L_{j}}$ the Josephson coupling
becomes very essential for the  calculation of $\eta$.  The solution
of this problem will be described in detail elsewhere.  The structure
of the effective core in this case becomes more complicated
\cite{feinberg}.  We may distinguish two contributions to the
viscosity coefficient ${(\eta=\eta_{2D}+\eta_{j})}$ which result from
the motion of 2D vortices ($\eta_{2D}$) and Josephson-like vortices
($\eta_{j}$) connecting 2D pancakes.  As for the term $\eta_{2D}$
 (this contribution will be the leading one at least for
${\bf j}\bot (xz)$), it has the form

\begin{equation}
\label{eta2D}
\eta_{2D}\simeq \eta_{0}
\left(\beta+
ln \left(\frac{1+2s R_{m}/\xi}{1+2s}\right)+
\left(1-\frac{1}{\sqrt{1+4s}}\right)
ln\frac{min[L_{j},L_{v}]}{R_{m}}\right)
\end{equation}
The logarithmic divergence of
$\eta_{2D}$ is cut off at a length scale $L_{j}$ if ${L_{j}<L_{v}}$.
For this angular interval the nonlinear effects in the vortex motion
appear for ${V>V_{c}\xi/L_{j}}$.
%%%%%%%%%%%%%%%%%%%%%%%%%%%%%%%%%%%%%%%%%%%%%%%%%%%%%%%%%%%%%%%%%%%

\section{CONCLUSIONS}

In summary, we have investigated theoretically some qualitatively new
features of 2D pancake vortex dynamics in multilayer superconducting
 structures.  The ohmic dissipation in nonsuperconducting layers
 results in the increasing of the viscosity coefficient.  In the limit
of zero interlayer Josephson coupling the viscosity coefficient for an
isolated 2D vortex appears to be logarithmically divergent.  This
divergence is cut-off at a length scale set by the vortex velocity V.
As a consequence the dynamic equation for an isolated  2D vortex
appears to be essentially nonlinear:  the viscosity coefficient depends
logarithmically on $V$. The main contribution to ohmic dissipation in
this case comes not from the core region
(${R\stackrel{_<}{_\sim}\xi}$) but from the large-$R$ domain
(${R\stackrel{_<}{_\sim}L_{v}}$). Therefore, for rather small
 distances between pancakes the dissipation produced by their motion
cannot be represented as a sum of contributions corresponding to
individual 2D vortices. This fact was demonstrated above for the
particular configuration of pancakes forming a tilted vortex line.

The equation of viscous motion for
a vortex line is obtained
and the effects of finite Josephson coupling are discussed.
 The viscosity coefficient is found to depend essentially on
the distance in the $x$-direction between pancakes and hence on the
tilting angle $\gamma$.
  For field orientations close to
the layers the nonlinear effects in the vortex motion are shown to
appear for ${V>V_{c}\xi/min[L_{j},Dtg\gamma]}$.
 These deviations from the ohmic behavior occur
even for slowly moving flux lines
when the transport current
is  much smaller than the Ginzburg-Landau critical current.  In this
nonlinear regime the velocity  dependence of $\eta$ coincides with
that  for an isolated 2D pancake.
Note that the dynamics of 2D pancakes can be strongly affected by the
interaction with defects of the crystalline structure, which we have
ignored in this paper. For 2D vortices localized by linear defects in
superconducting layers these phenomena were studied previously in
Ref.~\cite{mints}.
%%%%%%%%%%%%%%%%%%%%%%%%%%%%%%%%%%%%

\section*{ACKNOWLEDGMENTS}

This work has been made possible by a fellowship of Royal Swedish
Academy of Sciences and carried out under the research program of
the International Center for Fundumental Physics in Moscow (ICFPM).

\end{document}